\documentclass{aa} 
\usepackage{graphicx}
\usepackage{txfonts}
\usepackage{amsmath,amsfonts,amssymb} 
\usepackage{natbib}
\usepackage{multicol}
\usepackage{lipsum}
\usepackage{booktabs}
\usepackage{threeparttable}
\usepackage{multirow}

\bibpunct{(}{)}{;}{a}{}{,}
\usepackage[colorlinks=true,urlcolor=blue,linkcolor=blue,citecolor=blue]{hyperref}
\usepackage{color}

\begin{document}

\title{Reactive collision of electrons with CO$^+$ in cometary coma}
\authorrunning{Moulane et al. 2018}     
\titlerunning{Reactive collision of electrons with CO$^+$ in cometary coma}     
\author{
        Y. Moulane\inst{1,2,3}$^\dagger$, J. Zs. Mezei\inst{3,4,5,6}, V. Laporta\inst{3,7}, E. Jehin\inst{2}, Z. Benkhaldoun\inst{1} and I. F. Schneider \inst{3,6}
           }
        
\institute{Oukaimeden Observatory, High Energy Physics and Astrophysics Laboratory, Cadi Ayyad University, Morocco\\
        $^\dagger$\email{\color{blue}youssef.moulane@doct.uliege.be}
                \and
                Space sciences, Technologies \& Astrophysics Research (STAR) Institute, University of Liège, Belgium
                \and
                Laboratoire Ondes et Milieux Complexes, CNRS-UMR-6294, Universit\'e du Havre, France
                \and 
                Laboratoire des Sciences des Proc\'ed\'es et des Mat\'eriaux, CNRS-UPR-3407, Universit\'e Paris 13, France
                \and
                Institute for Nuclear Research, Hungarian Academy of Sciences, Debrecen, Hungary
                \and 
                Laboratoire Aim\'e Cotton, CNRS-UMR-9188, Universit\'e Paris Sud, ENS Cachen et Universit\'e Paris Saclay, France
                \and
                Department of Physics and Astronomy, University College London, London WC1E 6BT, UK\\
        }

\date{Received/accepted}

\abstract{In order to improve our understanding of the kinetics of the cometary coma, theoretical studies of the major reactive collisions in these environments are needed. Deep in the collisional coma, inelastic collisions between thermal electrons and molecular ions result in recombination and vibrational excitation, the rates of these processes being particularly elevated due to the high charged particle densities in the inner region.}
{This work addresses the dissociative recombination, vibrational excitation, and vibrational de-excitation of electrons with CO$^+$ molecular cations. The aim of this study is to understand the importance of these reactive collisions in producing carbon and oxygen atoms in cometary activity.}
{The cross-section calculations were based on Multichannel Quantum Defect Theory. The molecular data sets, used here to take into account the nuclear dynamics, were based on {\textit {ab initio}} R-matrix approach.}
{The cross sections for the dissociative recombination, vibrational excitation, and vibrational de-excitation processes, for the six lowest vibrational levels of CO$^+$ - relevant for the electronic temperatures observed in comets - are computed, as well as their corresponding Maxwell rate coefficients. Moreover, final state distributions for different dissociation pathways are presented.}   
{Among all reactive collisions taking place between low-energy electrons and CO$^+$, the dissociative recombination is the most important process at electronic temperatures characterizing the comets. We have shown that this process can be a major source of O($^3$P), O($^1$D), O($^1$S), C($^3$P) and C($^1$D) produced in the cometary coma at small cometocentric distances.}


\maketitle
        
\section{Introduction}

The carbon monoxide ion CO$^+$ is one of the most abundant ions detected in the interstellar medium~\citep{Erickson1981,Latter1993,Stoerzer1995,Fuente1997} and in the coma and tail region of comets, and it is of key relevance for the Martian atmosphere ~\citep{Fox1999}.

The cometary coma is the gaseous envelope around the comet nucleus, and consists of released molecules and dust grains created and dragged from the nucleus by solar heating and sublimation. Its shape can vary from one comet to another, while its formation depends on the comet's distance to the Sun and the relative amount of dust and produced gases. Due to the intense solar radiation, different light, and heavier ionising particles, it is the playground of physical and chemical processes involving various carbon-, oxygen-, hydrogen-, and nitrogen-based, neutral and/or charged, atomic and/or molecular  species.

\begin{table*}
\centering      
\caption{Relevant reactive processes producing CO$^+$ ions in the cometary coma.}  
\begin{tabular}{ll}
\hline  
\hline
Reaction        \hspace{5cm} & Reference \\
\hline           
h$\nu$ + CO $\longrightarrow$ CO$^+$ + e$^-$        & \cite{Ip1976}  \\
h$\nu$ + CO$_2$ $\longrightarrow$ CO$^+$+ O + e$^-$ & \cite{Huebner1980}   \\
e$^-$ + CO  $\longrightarrow$ CO$^+$ + 2e$^-$       & \cite{Vojnovic2013}  \\
e$^-$ + CO$_2$ $\longrightarrow$ CO$^+$ + O + 2e$^-$& \cite{Itikawa2002,Huebner1991} \\
H$_2$O$^+$ + CO $\longrightarrow$ CO$^+$ + H$_2$O   & \cite{Haider2005} \\
H$_2$CO$^+$ + CO $\longrightarrow$ CO$^+$ + H$_2$CO & \cite{Haider2005} \\
H$_2^+$ + CO $\longrightarrow$ CO$^+$ + H$_2$       & \cite{Kim1975} \\
H$^+$ + CO $\longrightarrow$ CO$^+$ + H             & \cite{Lopez2017,Tinck2016}\\
\hline
\hline

\end{tabular}

\label{tab1}
\end{table*}

The CO$^+$ ion, among many others, has been detected by several spacecraft missions to different comets, such as Giotto mission bond to the comet Halley, which detected it by ion mass spectroscopy, at a distance from approximately $1300$ km to about $7.5\times10^6$ km measured from the nucleus of the comet \citep{Balsiger1986,Huebner1991}. Moreover, the distribution of CO$^+$ on the surface and in the cometary coma of comet 29P/Schwassmann-Wachmann 1 was measured using spectroscopic observations \citep{Cochran1991}.
 
Carbon species can have different origins in cometary coma. Carbon monoxide is thought to originate
natively, from direct sublimation from the nucleus and from photo destruction of carbon dioxide, and in some cases from "extended sources", that is, involving production throughout the coma and not only at or near the surface of the nucleus, chemical reactions in the inner coma, or degradation of high molecular weight organic compounds present in cometary grains \citep{BockeleMorvan2010}.

The CO$^+$ ion temperature in cometary coma also varies from one comet to another, depending on the variation of heliocentric distance and on the solar wind, which is responsible for the comet's activity. 
For example, in coma of comet Halley, where the CO$^+$ ion was detected, the temperature of ions varies between $10^3$ K and $10^4$ K at distances of 5$\times$10$^3$ to 2.5$\times$10$^4$ km from the nucleus of the comet \citep{Balsiger1986}. The High Intensity Spectrometer (HIS) instrument of the ion mass spectrometer on board the Giotto spacecraft identified the contact surface at $4800$ km distance from the comet's nucleus. This boundary is clearly marked by a drastic drop in the temperatures of different ion species from about $2000$ K outside to values as low as $300$ K inside \citep{Schwenn1989}. \cite{Eberhardt1995} derived the electronic temperature profile using the H$_3$O$^+$/H$_2$O$^+$ ratios measured by the Giotto Neutral Mass Spectrometer in the inner coma of the comet P/Halley \citep{Krankowsky1986}. They also show increasing electronic temperature as a function of cometocentric distance. 

The different chemical processes that can produce CO$^+$ molecular ions in the inner coma are summarized in Table \ref{tab1}. All these processes contribute to cometary activity, which has its origin in the interaction of the solar wind with the comet nucleus \citep{Mendis1985,Broiles2015}, where the mother molecules can be found. The solar wind interacts very strongly with the extensive cometary coma, and the various interaction processes are initiated by the ionization of cometary neutrals \citep{Cravens1987}. The CO$^+$ ion is mainly produced by photoionization and electron impact ionization of CO and CO$_2$ molecules \citep{Ip1976,Huebner1980,Itikawa2002,Vojnovic2013}. The photoionization of these neutrals by the solar extreme ultraviolet radiation (EUV) has been assumed to be the main source of ionization near comets \citep{Mendis1985}, although both charge exchange with the solar wind protons and electron impact ionization have also been suggested \citep{Wallis1973,Kim1975,Kimura2000,Tinck2016,Lopez2017}. The charge exchange between CO and H$_2$O$^+$ at small distances and the reaction between CO and  H$_2$CO$^+$ at large distance from the cometary nucleus can contribute to CO$^+$ production \citep{Haider2005,Cessateur2016}.  
The high density of electrons and molecular ions in the cometary coma facilitates the reactive processes among them, such as the dissociative recombination (DR). This process plays an important role in producing numerous carbon and oxygen atoms in metastable states \citep{Raghuram2016,Decock2013,Bhardwaj2012}. Moreover, \cite{Feldman1978} showed that the DR of CO$^+$ by low-energy electron impact is the dominant source of carbon atoms rather than the photodissociation of CO in comets Kohoutek (1973 XII) and West (1976 VI). 

In the present work, we extend our previous study on the reactive collisions between electrons and carbon monoxide cations \citep{Mezei2015} to more excited states. The major one is the dissociative recombination, which takes place \textit{via}  two mechanisms: (i) the {\it direct} process, consisting in the capture into a dissociative state of the neutral system, CO$^{**}$:
\begin{equation}
\mbox{CO}^{+}(v_{i}^{+}) + e^{-}\rightarrow \mbox{CO}^{**} \rightarrow 
\mbox{C} + \mbox{O}\,;
\label{eq:dra}
\end{equation}
\noindent and (ii) the {\it indirect} process, where the capture occurs into a Rydberg state of the neutral molecule CO$^{*}$, subsequently  predissociated by the CO$^{**}$ state,
\begin{equation}
\mbox{CO}^{+}(v_{i}^{+}) + e^{-}\rightarrow \mbox{CO}^{*} \rightarrow
\mbox{CO}^{**}  \rightarrow  \mbox{C} + \mbox{O}\,.
\label{eq:drb}
\end{equation}
We note here that we follow the standard nomenclature in this work, namely that
CO$^{**}$ and CO$^{*}$ represent the doubly excited and singly excited
states of CO, respectively, \citep{Mezei2015}. 

Meanwhile, the competitive processes with respect to DR are:
\begin{equation}
\mbox{CO}^+(v_i^+) + e^- \longrightarrow \mbox{CO}^+(v_f^+)  + e^-\,,
\label{eq:drc}
\end{equation}
\noindent 
that is, the elastic collisions (EC: $v_i^+ = v_f^+$), the
vibrational excitation (VE: $v_i^+ < v_f^+$), and  the vibrational de-excitation (VdE: $v_i^+ > v_f^+$), 
where $v_i^+$ and $v_f^+$ stand for the initial and final vibrational quantum numbers of the target ion, respectively, and rotational structure is neglected.

In order to understand the role of this molecular ion in the coma and in the tail regions of comets, considerable effort has been directed towards modeling its chemistry \citep{Huebner1980}.
In this context, the DR with electrons is a major destruction mechanism of CO$^+$, and it is also believed to be responsible for the C($^1$D) emissions observed in comet spectra \citep{Feldman1978} through the reactions:
\begin{align}
\mbox{CO}^+ + e^-               & \longrightarrow \mbox{C}(^3\mbox{P}) + \mbox{O}(^3\mbox{P})\,,\\
                                & \longrightarrow \mbox{C}(^1\mbox{D}) + \mbox{O}(^3\mbox{P})\,,\\
                                & \longrightarrow \mbox{C}(^3\mbox{P}) + \mbox{O}(^1\mbox{D})\,.
\end{align}

The main goal of this work is to evaluate all the DR, VE, and VdE cross sections and rate coefficients of CO$^+$ relevant for cometary comae. This was performed by extending our previous nuclear
dynamics computations to high vibrational levels of the target. The Multichannel Quantum Defect Theory (MQDT) is applied to these computations starting from molecular data sets provided in \cite{Mezei2015} based on R-matrix 
calculations of \cite{Chakrabarti2006,Chakrabarti2007}. 

The paper is organized as follows: A brief description of molecular data sets and of the MQDT approach used in the calculations is given in Section \ref{mqdt}. In Section \ref{results}, the obtained cross sections and rate coefficients are presented, followed by their interpretation and discussion.  Finally, Section \ref{conclusion} contains our conclusions and future plans.

\section{Theoretical approach: the main steps and ideas}
\label{mqdt}

The quantum defect theory was originally introduced in the field of atomic physics by \citep{Seaton1966} in order to describe the optical transitions in alkali atoms. By introducing the frame transformation, the method was extended by \cite{Fano1970} to treat the coupling between the motion of the electrons and the rotation of the nuclei in a molecule. This approach was generalized by \cite{Jungen1977} and \cite{Greene1985} to treat ro-vibronic couplings in diatomic molecules, restricting however to ionization channels. The inclusion of the dissociation channels in this theory by \cite{Giusti1980},  improving the early work of  \cite{Lee1973},  opened the way to the most accurate treatment of the  DR.

\begin{table}[t]
        \centering      
        \caption{Vibrational energy levels of CO$^+$ with respect to the vibrational ground state.}              
        \begin{tabular}{cccccc}
                \hline  
                \hline
        $\nu_i^+$ & $\varepsilon_{v_i^+}$ (eV) & $\nu_i^+$ & $\varepsilon_{v_i^+}$(eV) & $\nu_i^+$ & $\varepsilon_{v_i^+}$(eV) \\
                \hline           
0&0.000  &18&4.274 &36&7.373\\
1&0.267  &19&4.479 &37&7.494\\
2&0.530  &20&4.680 &38&7.609\\
3&0.789  &21&4.877 &39&7.715\\ 
4&1.045  &22&5.070 &40&7.814\\
5&1.298  &23&5.259 &41&7.904\\ 
6&1.547  &24&5.445 &42&7.987\\
7&1.793  &25&5.627 &43&8.061\\
8&2.035  &26&5.806 &44&8.127\\ 
9&2.273  &27&5.982 &45&8.185\\
10&2.509 &28&6.156 &46&8.236\\ 
11&2.741 &29&6.326 &47&8.279\\ 
12&2.970 &30&6.492 &48&8.316\\ 
13&3.195 &31&6.655 &49&8.348\\
14&3.418 &32&6.812 &50&8.373\\ 
15&3.637 &33&6.964 &51&8.393\\
16&3.853 &34&7.108 &52&8.408\\
17&4.065 &35&7.244 & & \\ 
                \hline
                \hline
                
        \end{tabular}
        \label{levels}
\end{table}

Since then, it was applied with great success to several diatomic systems like H$_2^+$ and its isotopologs in a broad range of energy \citep{Giusti1983,Schneider1991,Takagi1993,Tanabe1995,Schneider1997,Amitay1999,Chakrabarti2013}, or systems of interest in planetary atmospheres like CO$^+$~\citep{Mezei2015} or
N$_2^+$ \citep{Little2014}, for molecules observed in the interstellar medium like CH$^+$ \citep{faure2017} and SH$^+$~\citep{kashinski2017}, and finally for molecular systems of interest for technological plasmas, such as BeH$^+$~\citep{niyonzima2017} and BF$^+$~\citep{Mezei2016}. It was also applied to triatomic systems such as H$_3^+$ \citep{Schneider2000,Orel2000,Kokoouline2005}.

The MQDT treatment of DR reactions (\ref{eq:dra}) and (\ref{eq:drb}) and its competitive processes (\ref{eq:drc}) rely on the knowledge of the potential energy curves (PECs) of the ground and relevant excited electronic states of the molecular cation, the PECs of the relevant "doubly excited" dissociating autoionizing states of the neutral molecule, as well as the PECs of "mono-excited" bound series of Rydberg states related to each electronic state of the target cation. These latter two sets of PECs allow us to define the quantum defects. In addition to the PECs, a set of electronic couplings between the two major fragmentation channels (ionization and dissociation) are needed to perform the nuclear dynamics.

A dissociation channel describing the atom-atom scattering consists in an electronically bound state whose potential energy in the asymptotic limit is situated below the total energy of the system. 

An ionization channel accounting for the electron-molecular ion scattering consists of a Rydberg series of excited states extrapolated above the continuum threshold - a vibrational level v$^+$ of the molecular ion, given in table~\ref{levels} with respect to the ground vibrational level - and an incoming/outgoing electron with angular momentum $l$. It is {\it `open'} if its corresponding threshold is situated below the total energy of the system, and {\it `closed'} otherwise. The existence of these latter channels is related to the {\it `indirect'} (\ref{eq:drb}) mechanism, and  its quantum interference with the {\it `direct'} (\ref{eq:dra}) one results in the {\it total} cross section. We have adopted a second-order perturbative solution of the Lippmann-Schwinger equation providing the reaction matrix, which accounts for all the major mechanisms driving the reactive collisions. This solution is exact if we neglect the energy dependence of the autoionization widths of the dissociative states.

Once the scattering matrix $S$ for the DR and vibrational transitions are determined, the respective global cross sections, as a function of the incident electron kinetic energy $\varepsilon$, are obtained by summation over all the dissociative states $d_j$ corresponding to the
relevant symmetries and to the projection of the total electronic angular momentum on the nuclear axes $\Lambda$ of the system:
\begin{equation}
\sigma_{diss \leftarrow v_i^+}(\varepsilon)=\dfrac{\pi}{4\varepsilon}\sum_{\Lambda,sym}\rho^{sym,\Lambda}\sum_{j,l}{|S_{d_{j},lv_{i}^+}|}^2,
\end{equation}
\begin{equation}
\sigma_{v_f^{+} \leftarrow v_i^+}=\dfrac{\pi}{4\varepsilon}\sum_{\Lambda,sym}\rho^{sym,\Lambda}\sum_{l',l} {|S_{l'v_f^{+},lv_{i}^+}-\delta_{l'l}\delta{v^+_fv_i^+}|}^2.
\end{equation}
Here $\rho^{sym,\Lambda}$ is the ratio of the multiplicities of the neutral system and the ion.
\begin{figure*}

        \resizebox{\hsize}{!}{
                \includegraphics{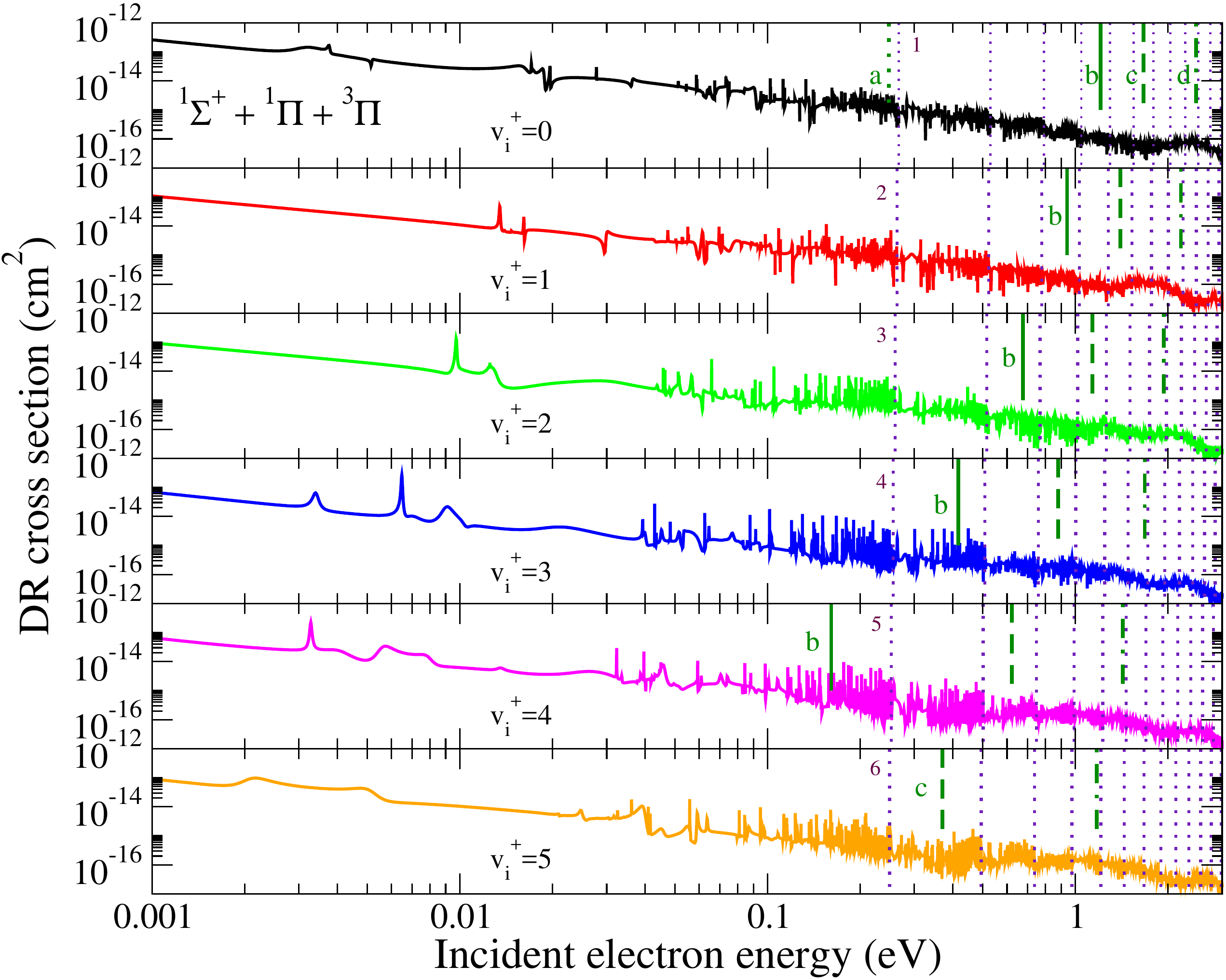}}
        \caption{Dissociative recombination  of CO$^+$ on its six lowest vibrational levels ($v_i^+$=0, 1, 2, 3, 4 and 5):  cross sections summed-up over all the relevant symmetries - see ~\cite{Mezei2015}. The dotted vertical indigo lines are the different ionization thresholds given by the vibrational levels of the molecular ion. The first ionization thresholds are indicated on the figures. The dark-green shorter vertical lines stand for the different dissociation limits measured from the initial vibrational levels of the ion, as follows: the dotted line (a) stand for the C($^1$D)$+$O($^1$D) limit, the solid lines (b) for the C($^3$P)$+$O($^1$S) one, the dashed lines (c) for C($^1$S)$+$O($^1$D) , and the dashed-dotted lines (d) for  C($^1$D)$+$O($^1$S).
        }       
        \label{cross}
\end{figure*}

\section{Results and Discussion}
\label{results}
In order to explore the reactive collisions of electrons with
$^{12}$C$^{16}$O$^+$
- the most abundant isotopolog of carbon monoxide -
in plasmas characterized by temperatures 
up to $5000$ K, we have extended the previous calculations of~\cite{Mezei2015} up to the fifth vibrational level of the target, $v_i^+=5$.

We have relied on the same molecular structure data, corresponding to the three most important symmetries contributing to these processes, namely $^1\Sigma^+$, $^1\Pi$ and $^3\Pi$, considering  four dissociative states for each symmetry.  For each available dissociative channel, we have considered its interaction with the most relevant series of Rydberg states, that is,  
 $s$, $p$, $d$ and $f$,
for the $^1\Sigma^+$ symmetry, 
and
$s$, $p$ and $d$
for the  $^1\Pi$ and $^3\Pi$ symmetries.

\subsection{Cross sections}
We have performed  calculations for incident electron energy in the range 0.01 meV - 3 eV.

\begin{table*}
        \begin{center}
                \caption{Dissociative recombination branching ratios as a function of the electron 
                 temperature, for the six lowest vibrational levels of CO$^+$.}
                \resizebox{\textwidth}{!}{%
                        \begin{tabular}{c|cccccc|cccccc|cccccc}
                                \hline
                                \hline
                                Dissociation   & \multicolumn{18}{c}{Electron temperature (K)} \\
                                Path & \multicolumn{6}{c|}{300} & &&\multicolumn{2}{c}{1000}  & && \multicolumn{6}{c}{5000} \\
                                \cline{2-19}
& v$_i^+$=0 & v$_i^+$=1 &v$_i^+$=2 &v$_i^+$=3 &v$_i^+$=4 &v$_i^+$=5  & v$_i^+$=0 & v$_i^+$=1 &v$_i^+$=2 &v$_i^+$=3 &v$_i^+$=4 &v$_i^+$=5 & v$_i^+$=0 & v$_i^+$=1 &v$_i^+$=2 &v$_i^+$=3 &v$_i^+$=4 &v$_i^+$=5  \\ 
                                \hline  
C($^3$P)+O($^3$P)&88.4\%&97\%&61\%&50.1\%&66.77\%&79.97\%&87.6\%&87.7\%&60.5\%&47.67\%&57.88\%&74.84\%&76.8\%&79.2\%&67.45\%&60.31\%&57.44\% &59.12\%\\
C($^1$D)+O($^3$P)&11.6\%&0.5\%&15\%&16.9\%& 4.08\%&1.53\%&12.1\%&0.7\%&12.5\%&15.74\%&4.70\%&2.14\%&12.5\%&3.1\%&5.7\%& 9.88\%&6.94\% &4.16\%\\
C($^1$D)+O($^1$D)&      &2.5\%&24\%&33\%  &29.15\%&18.50\% &0.1\%&11.6\%&27\%&36.58\%&37.40\%&22.76\% &10\%&16.2\%&24.45\%&25.60\%&28.82\%&24.68\%\\
C($^1$S)+O($^1$D)&      &     &    &      &       &   &      &      &      &        &0.01\% &0.16\% &0.4\%&1\% &1.6\%&2.82\% &4.54\% &8.18\% \\
C($^1$S)+O($^3$P)& --   &--   &--  & --   & --    & --&--    &--    &   -- & --     &  --   &0.09\% &0.2\%&0.4\% &0.7\%&1.21\% &0.25\% &2.85\% \\
C($^3$P)+O($^1$S)&      &     &    &      &       &   &      &      &      &        &       & &0.02\% &0.1\%&0.04\% &0.08\% &1.85\% &0.76\% \\
C($^1$D)+O($^1$S)&      &     &    &      &       &   &      &      &      &        &       & &       &      &0.04\% &0.08\% &0.15\% &0.23\% \\
                                \hline
                                \hline
                        \end{tabular}}
                        \label{tab2}
                \end{center}
        \end{table*}

The DR cross sections are displayed in Figure~\ref{cross}. They are characterized by resonance structures due to the temporary captures into vibrational levels of Rydberg states embedded in the ionization continuum ({\it closed} channels, {\it indirect} process), superimposed on a smooth background originating in the  {\it direct} process~\citep{Mezei2015}.

The ionization thresholds (vibrational levels of the molecular ion: Table~\ref{levels}) shown as dotted vertical lines in Fig.~\ref{cross} act as accumulation points for these Rydberg resonances. Moreover, the asymptotic limits of the dissociation channels opening progressively are shown with shorter dark-green vertical lines in Fig.~\ref{cross}, corresponding to the atomic pairs of states C($^1$D)$+$O($^1$D), C($^3$P)$+$O($^1$S) C($^1$S)$+$O($^1$D) and C($^1$D)$+$O($^1$S). We notice that the C($^3$P)$+$O($^3$P) and C($^1$D)$+$O($^3$P) limits are open at zero collision energy. 

\begin{figure*}
	\centering
	\resizebox{\hsize}{!}{
		\includegraphics{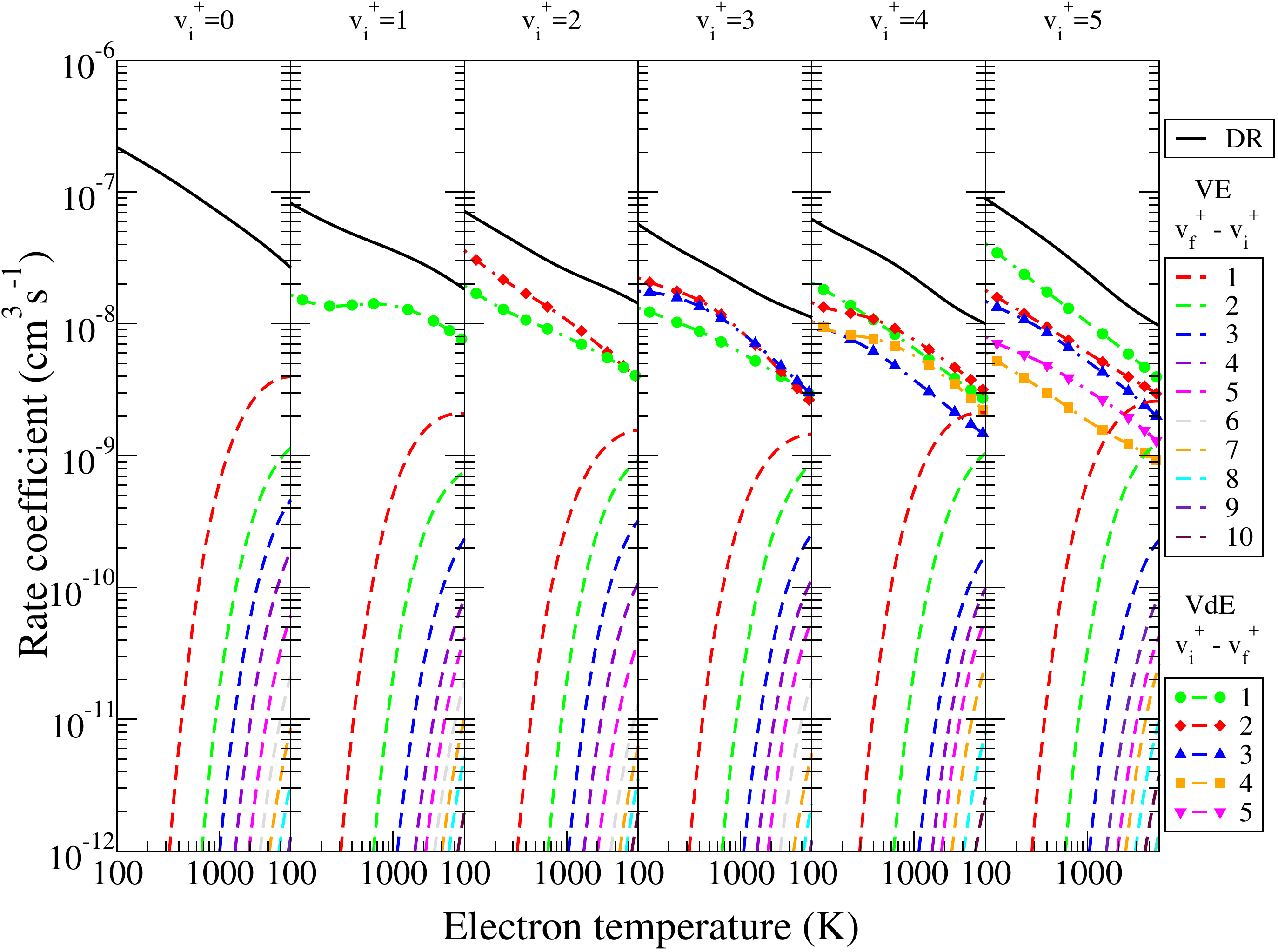}}
	\caption{Dissociative recombination (DR), vibrational excitation (VE) and vibrational de-excitation (VdE) of CO$^+$ on its lowest six  
		vibrational levels ($v_i^+$=0, 1, 2, 3, 4 and 5): 
		rate coefficients as a function of the electron temperature.}
	\label{fig:rate}
\end{figure*}

The role of each dissociation channel in the total cross section depends on the strength of the valence-Rydberg electronic couplings and on the position of the point of crossing between the PEC of the dissociative valence state and that of the target ion.

Figure~\ref{cross} shows that the cross section for the ground vibrational level has the largest cross section: about four times larger below $700$ meV and above $2$ eV, while in-between, the maximum deviation among all the cross sections is smaller than a factor of two. At low-energy collision range, as the vibrational quantum of the initial ionic target is increased, one can observe a systematic decrease of the total cross section,  except for v$_i^+=5$. In this latter case, the PECs of the open valance states  of $^1\Sigma^+$ symmetry correlating to the C($^3$P)$+$O($^3$P) and C($^1$D)$+$O($^3$P) atomic limits  both have favorable crossings with the ion PEC, this symmetry displaying the largest  valence-Rydberg electronic couplings (see Fig. 2 from \cite{Mezei2015}), leading to an increase in the cross section.

Another interesting feature can be observed in the high-energy range. One can observe a revival in the cross section, which is due to the opening of the dissociation states correlating to the C($^1$S)+O($^1$D) and C($^1$D)+O($^1$S) atomic limits represented by dashed and dotted-dashed vertical dark-green lines in Fig.~\ref{cross}. The maximum in the cross section can be observed at the collision energies where the crossings of these newly open dissociative states with the ion's ground electronic state become favorable.

Besides the total cross sections, an important characteristic of the collision is the branching ratios;  
Table \ref{tab2} shows them, the results being obtained after  summing the  three  relevant  symmetries $^1\Pi$, $^1\Sigma^+$ and $^3\Pi$ . This gives an estimation of the atomic neutral species that are formed in the DR of CO$^+ (v_i^+=0\to5)$. At low collision energy and/or electron temperature, the dominant dissociation pathway for the DR is the one that correlates with the C($^3$P)+O($^3$P) atomic limit. In this energy range we have obtained good agreement with experimental measurements of \cite{Rosen1998}.

The main channels for producing various atomic/neutral species such as O($^3$P), O($^1$D), O($^1$S), C($^3$P), and C($^1$D) in the inner cometary coma \citep{Raghuram2016} are the dissociative excitation of the neutral molecular species by photons and supra-thermal electrons such as photoelectrons, as well as the DR of the molecular ions. For example, the oxygen atoms (O($^1$D) and O($^1$S)) are produced by the photodissociation of CO, CO$_2$ and H$_2$O molecules coming from the sublimation of the cometary ices \citep{Bhardwaj2012,Decock2013,Raghuram2014,Decock2015}. Meanwhile, using a coupled-chemistry-emission model, 
\cite{Raghuram2016} showed that the DR of the CO$^+$ ion is an important source of C($^1$D). Table \ref{tab2} also shows that the DR of the CO$^+$ ion is one of the main sources of various metastable species, excited states that could not be formed by optical transitions in the inner coma at low energy. For example, the DR of CO$^+(v_i^+=1)$ is the major source of C($^3$P) and O($^3$P). 

One can conclude that the DR process plays an important role in producing the carbon and oxygen atoms in metastable excited states at small cometocentric distances, where the electronic temperature is very low \citep{Krankowsky1986,Eberhardt1995}. Generally speaking, the DR tends to be more important on the tail-ward sides of comets, since ions are channeled on the tail axis by the low magnetic pressure in the magnetic neutral sheet \citep{Gombosi1996}, which leads to high electron and ion density there. Furthermore, the DR is enhanced in cometary tails by the low temperature prevailing there, which elevates the rate constants of this process \citep{Haberli1973}.

\subsection{Rate Coefficients}

In order to contribute to the modeling of the cometary coma, we have computed the rate coefficients for DR, VE, and VdE, starting from the previously produced cross sections, and assuming that the velocity/kinetic energy distribution of the electrons is Maxwellian:

\begin{equation}
\alpha(T)=\int\dfrac{8 \pi m_{e}\varepsilon}{(2 \pi m_{e}kT)^{3/2}}\sigma(\varepsilon) e^{-\frac{\varepsilon}{kT}}d\varepsilon,
\label{eq:rate}
\end{equation}
\noindent where
$m_e$ is the mass of the electron, 
$k$ is the Boltzmann constant and $T$ stands for the electron's temperature.

Figure \ref{fig:rate} shows the DR (solid black curve), VE (colored dashed curves), and VdE (colored dashed-dotted curves with symbols) rate coefficients for the six lowest vibrational levels of the CO$^+$ molecular cation as a function of the electronic temperature.
 
The highest rates for DR and VE correspond to a vibrationally relaxed target, a decrease from the first to the fourth excited state, and an increase for the fifth one, whereas the VdE becomes progressively more important when the excitation of the target increases.

In the inner coma of a comet, the molecules tend to be in thermodynamic equilibrium, rather than in a fluorescence one. When they drift outwards, one can find regions where these species will be alternatively in one type of these equilibria. Considering the cometary coma in a thermal equilibrium condition, the DR rate coefficients of CO$^+$ decreases when the temperature increases. 
\begin{figure}[t]
	\centering
	\includegraphics[scale=0.42]{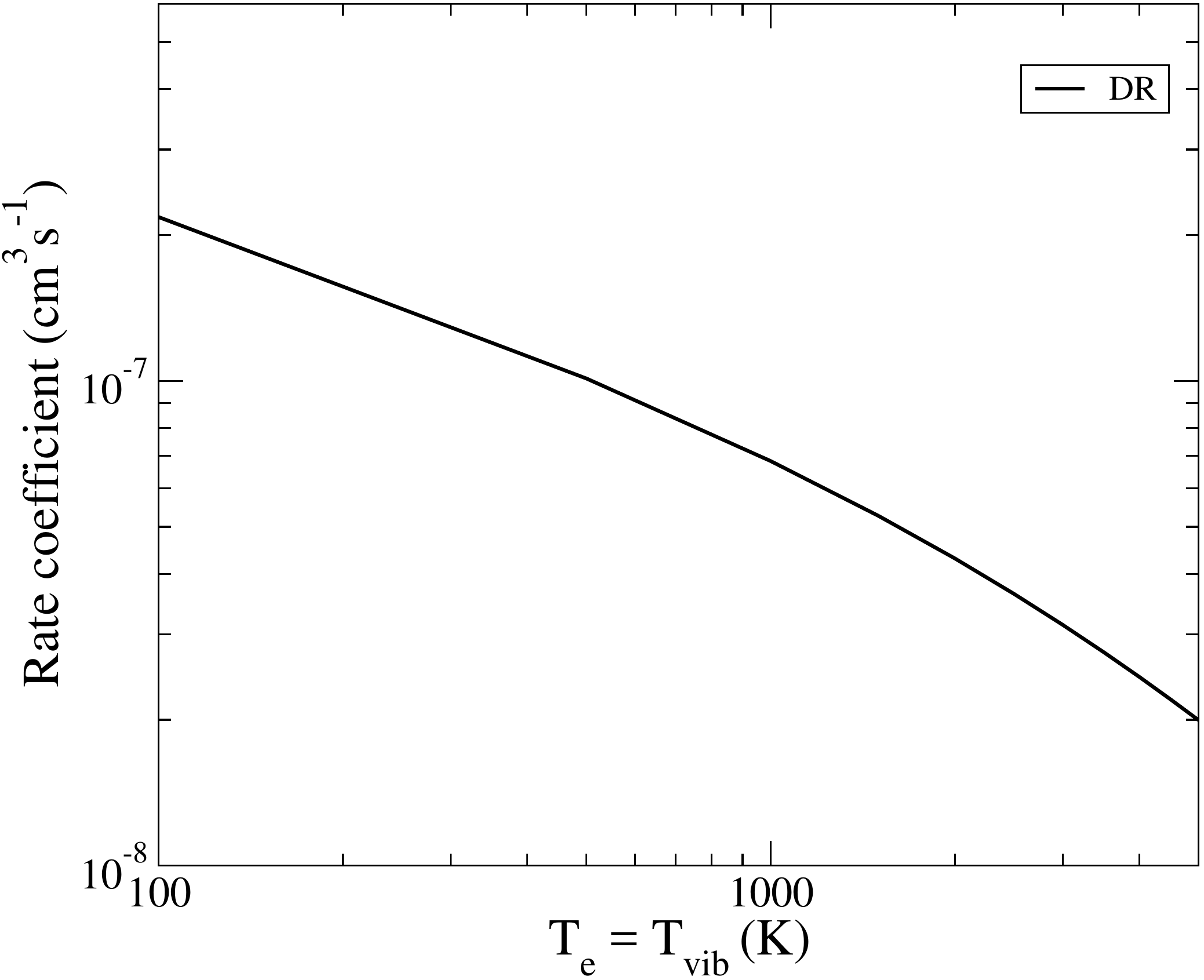}  
	\caption{Boltzmann vibrational average dissociative recombination of CO$^+$: rate coefficients as functions of the electron temperature, considered equal to the vibrational temperature.}
	\label{te_tvib}
\end{figure}

\begin{figure}[t]
	\includegraphics[scale=0.42]{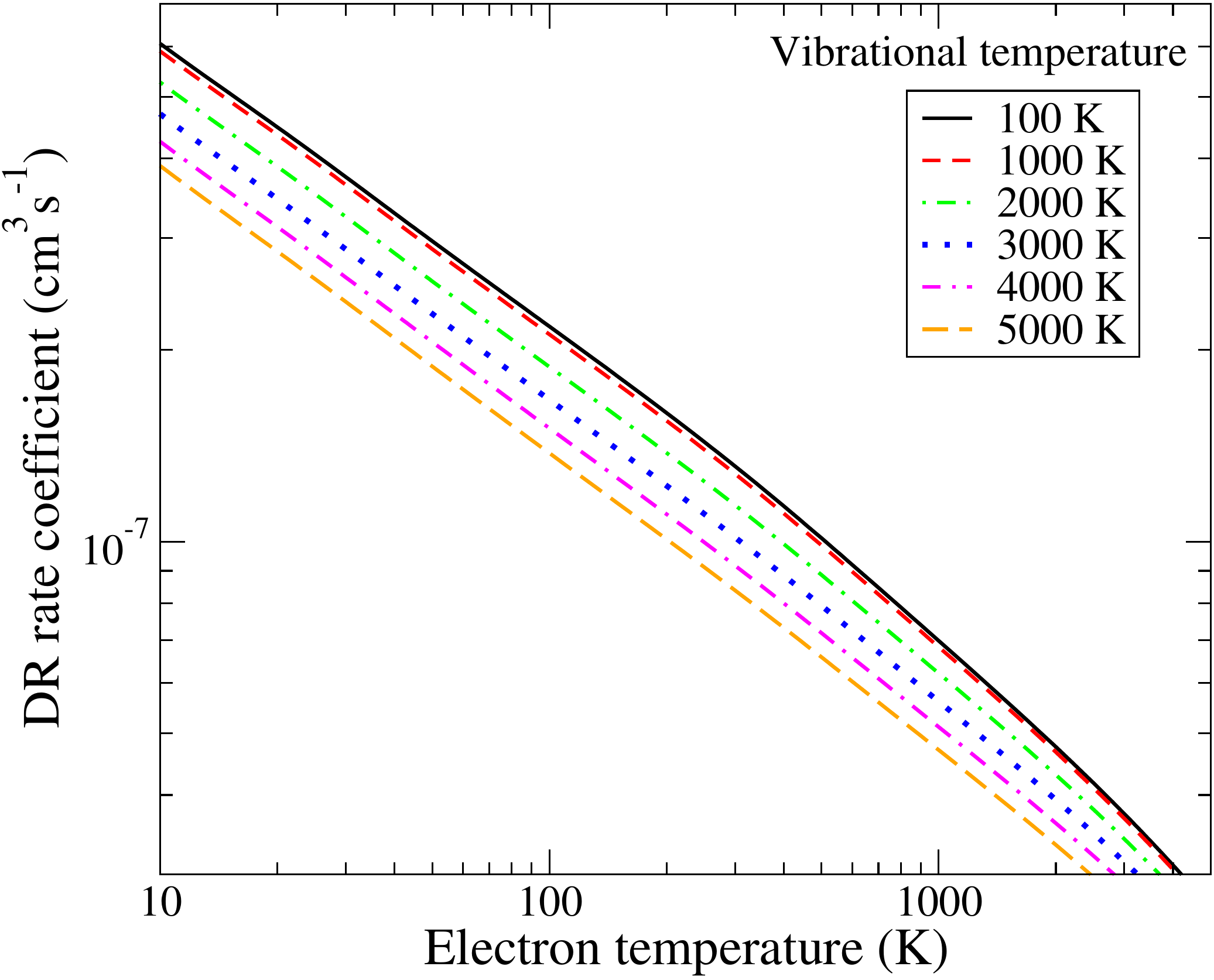}   
	\caption{Boltzmann vibrational average dissociative recombination of CO$^+$: rate coefficients as functions of the electron temperature  and of the vibrational temperature.}
	\label{temp}
\end{figure}

\begin{figure}[t]
	\includegraphics[scale=0.42]{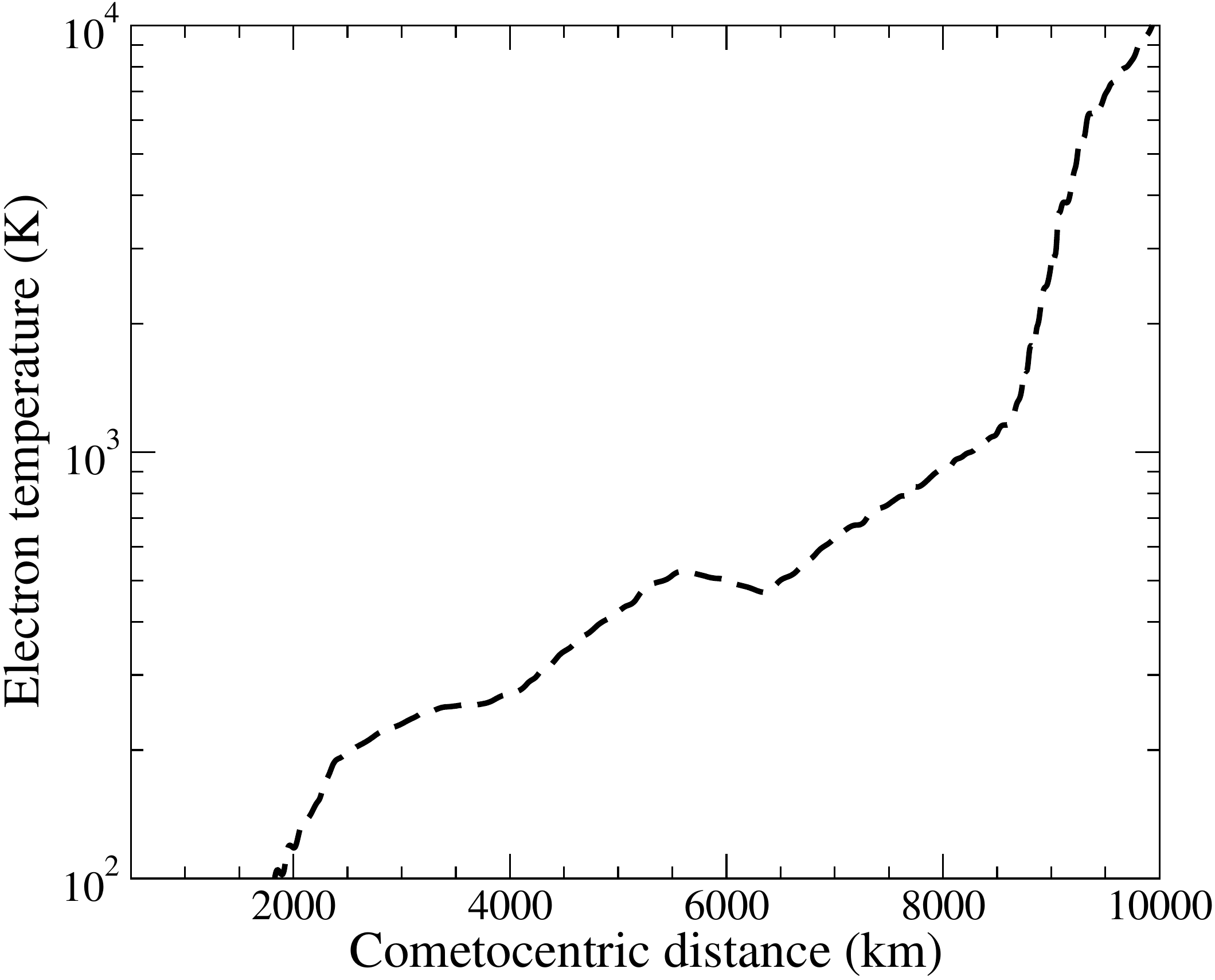} 
	\caption{Electron temperature profile as a function of the cometocentric distance \citep{Eberhardt1995,Gombosi1996}.}
	\label{profile}
\end{figure}

\begin{figure}[t]
	\includegraphics[scale=0.4]{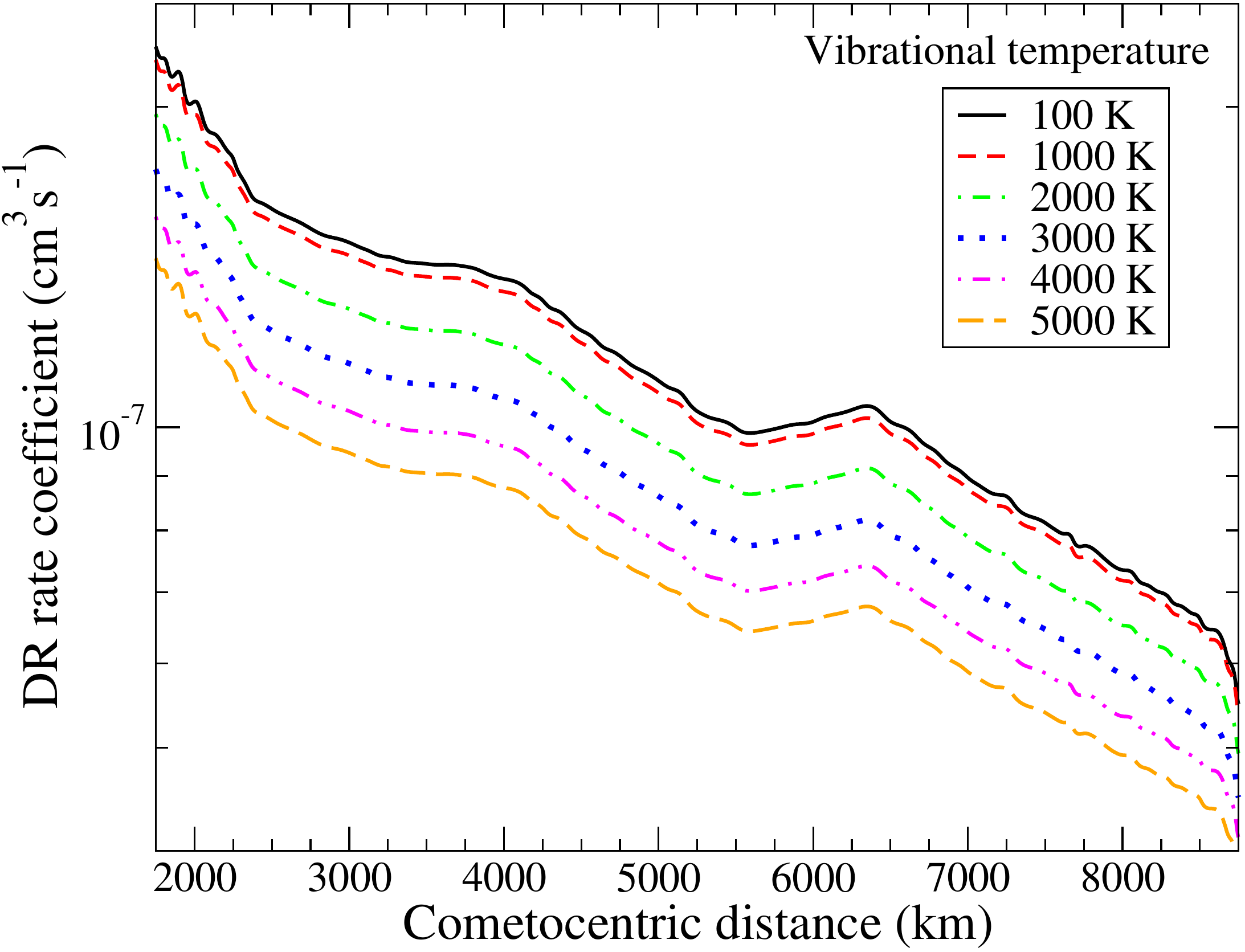}        
	\caption{Boltzmann vibrational average dissociative recombination of CO$^+$: rate coefficients as functions of the cometocentric distance  and of the vibrational temperature.}
	\label{cometocentric}
\end{figure}

Figure \ref{te_tvib} shows the average DR rate coefficient of all vibrational levels considered in our calculation at thermodynamical equilibrium, when the vibrational temperature is equal to the electronic one ($T_e=T_{\text{vib}}$). In addition to this, Fig. \ref{temp} shows the same DR rate coefficients with the vibrational levels considered according to a Maxwell distribution, for a wide range of vibrational temperatures, from $100$ to $5000$ K.

\subsection{Rate coefficients as a function of cometocentric distance}

In order to express our rate coefficients as functions of the cometocentric distance, rather than of the electronic temperature, we used the temperature profile resulting from the observations of the Giotto Neutral Mass Spectrometer at Halley's coma \citep{Eberhardt1995,Gombosi1996} - Figure \ref{profile}.

The steep increase of the electron temperature as a function of the cometocentric distance is consistent with earlier theoretical calculations \citep{Ip1985,Korosmezey1987,Marconi1988,Gan1990,Huebner1991}, and with the more comprehensive treatment of the electron temperature given by \citep{Haberli1996}.

Corroborating the data of Figs. \ref{temp} and \ref{profile} results in  Fig. \ref{cometocentric}, which shows the variation of the average DR rate coefficient as a function of the cometocentric distance for different vibrational temperatures  (from $100$ K to $5000$ K). These results suggest that the DR rate coefficients are very important at small distances from the nucleus of the comet and at low vibrational temperatures ( up to $\sim$ $1000$ K) \citep{Eberhardt1995,Gombosi1996}. Furthermore, one can conclude that the DR is among the most important collision processes in cometary coma at small cometocentric distances. At large distances, where the electronic temperature is much higher in absolute value, the DR is less important, but also in comparison with increasingly fast competitive processes like VE or dissociative excitation.

\section{Conclusion}
\label{conclusion}

The present theoretical results (Figs. \ref{cross} and \ref{fig:rate}) provide the most complete low-energy collisional data on electron-induced dissociative recombination, vibrational excitation, and de-excitation of the CO$^+$ molecular cation based on potential energy curves and electronic couplings calculated, derived, and calibrated from {\it ab initio} R-matrix calculations and spectroscopical data. Cross-sections between $0.01$ meV and $3$ eV, and Maxwell rate coefficients between $100$ and $5000$ K were calculated for DR, VE, and VdE of electrons with CO$^+$(X$^2\Sigma^+$) ions in their six lowest vibrational levels. 

We have focused on the important role of dissociative recombination in producing various atomic species in metastable excited states at small cometocentric distances (Table \ref{tab2}). According to our calculations, the DR of CO$^+$ may be the major source of metastable O($^1$S) and O($^1$D) oxygen atoms responsible for the green (5577 \AA) and red doublet (6300, 6364 \AA) emission lines observed  in the cometary coma \citep{Bhardwaj2012,Raghuram2014}. Moreover, the dissociative recombination process can be considered as a source of excited C($^1$D) and C($^3$P) atoms, whose emission has been detected in the Hale-Bopp comet \citep{Feldman1978,Raghuram2016}.

Using different vibrational temperatures of the molecular cation target, we have calculated the DR rate coefficients as a function of cometocentric distance (Figure  \ref{cometocentric}), pointing out  the importance of the DR process in the nucleus of the comet. The electron temperature profile used in our calculation is based on the measurements of Giotto with the neutral mass spectrometer at Halley's coma \citep{Eberhardt1995,Gombosi1996}.

The next step in our study of the relevance of collisional processes in comets consists in  performing calculations on polyatomic systems such as  H$_3$O$^+$ and H$_2$O$^+$, because of their significant abundance in the cometary coma \citep{Haider2005,Morvan2010,Vigren2013,Fuselier2016,Vigren2016}.

\section*{Acknowledgments}
This work is supported by BATTUTA Project (Building Academic Ties Towards Universities through Training Activities) in the frame of the Erasmus Mundus program, at LOMC UMR-CNRS-6294 of Le Havre University. YM thanks the SRI department, especially Mrs. Martine Currie, for outstanding hospitality.
The authors acknowledge support from the IAEA \textit{via} the Coordinated Research Project ``Light Element Atom, Molecule and Radical Behaviour in the Divertor and Edge Plasma Regions", from Agence Nationale de la Recherche via the projects `SUMOSTAI' (ANR-09-BLAN-020901) and  `HYDRIDES' (ANR-12-BS05-0011-01), from the IFRAF-Triangle de la Physique via the project `SpecoRyd', and from the CNRS \textit{via} the programs `Physique et Chimie du Milieu Interstellaire', and the  PEPS projects `Physique th\'{e}orique et ses interfaces' TheMS and TPCECAM. They also thank for generous financial support from La R\'egion Haute-Normandie \textit{via}  the GRR Electronique, Energie et Mat\'eriaux, from the ``F\'ed\'eration de Recherche Energie, Propulsion, Environnement", and from the LabEx EMC$^3$ and FEDER via the projects PicoLIBS (ANR-10-LABEX-09-01), EMoPlaF and CO$_2$-VIRIDIS. IFS also thanks the Laboratoire  Aim\'e Cotton for hospitality. JZM acknowledges support from USPC \textit{via} ENUMPP and Labex SEAM.

\bibliographystyle{aa}
\bibliography{aa.bib}

\end{document}